# CKFNet: Neural Network Aided Cubature Kalman filtering

Jinhui Hu, Haiquan Zhao, *Senior Member, IEEE*, and Yi Peng

*Abstract*—The cubature Kalman filter (CKF), while theoretically rigorous for nonlinear estimation, often suffers performance degradation due to model-environment mismatches in practice. To address this limitation, we propose CKFNet-a hybrid architecture that synergistically integrates recurrent neural networks (RNN) with the CKF framework while preserving its cubature principles. Unlike conventional model-driven approaches, CKFNet embeds RNN modules in the prediction phase to dynamically adapt to unmodeled uncertainties, effectively reducing cumulative error propagation through temporal noise correlation learning. Crucially, the architecture maintains CKF's analytical interpretability via constrained optimization of cubature point distributions. Numerical simulation experiments have confirmed that our proposed CKFNet exhibits superior accuracy and robustness compared to conventional model-based methods and existing KalmanNet algorithms.

## I. INTRODUCTION

ACHIEVING accurate state estimation is a significant challenge across diverse domains, including power systems [1], navigation systems [2], and the Internet of Things [3]. Among the classical state estimation algorithms, the Kalman filter (KF) [4] and the least squares state estimator stand out. To address the challenge of extending the Kalman Filter (KF) to effectively handle nonlinear system state estimation problems, several nonlinear variants of the KF have been proposed [5]-[8]. Where the Cubature Kalman Filter (CKF) outperforms the Extended Kalman Filter (EKF) by leveraging the cubature rule to accurately model nonlinear dynamics, circumventing the first-order linearization approximations inherent to EKF. However, the KF and its variants are predicated on an accurate representation of the underlying dynamical system, and they often experience performance degradation when faced with the uncertainties inherent in real-world scenarios [9]-[11].

Consequently, existing research on the KF primarily addresses these types of challenges. Among them, adaptive Kalman, variational Bayesian Kalman filtering, and robust Kalman filtering have been studied in different forms for this problem [12]-[16]. However, these variants struggle with real-time, precise noise estimation, which limits their performance in dynamic environments. Conversely, deep learning offers the potential for temporal feature extraction; however, its data requirements and opacity hinder its deployment in low-data or safety-critical contexts [17]-[26]. Hybrid frameworks, such as KalmanNet [9][27]-[29], have been developed to address this issue by integrating neural networks with Kalman filters to achieve a balance between adaptability and interpretability. However, these frameworks still face the challenge of prediction-phase optimization, which can lead to the accumulation of errors. Classical nonlinear filters mitigate this via deterministic state distribution modelling. A potential avenue for enhancing the synergy between data-driven flexibility and model-based error suppression could be found in the integration of prediction-phase learning, akin to that employed in CKF/UKF's sampling, into neural filters. This approach has the potential to facilitate robust, interpretable estimation in scenarios characterized by uncertainty and limited data availability.

To address the above problems, we propose CKFNet, a novel neural-enhanced filter that preserves the interpretability of the Cubature Kalman Filter (CKF) while addressing model uncertainties through neural networks. Key innovations include: 1) Prediction-phase optimization: Neural networks learn the distribution and weights of cubature points, reducing cumulative errors in real-time estimation by aligning propagated states with ideal distributions; 2)Update-phase adaptation: Leveraging cubature point properties as network inputs, CKFNet dynamically learns noise statistics and directly computes the Kalman gain, bypassing linearized approximations used in KalmanNet; 3)Hybrid architecture: Combines CKF's deterministic cubature rules with data-driven noise modeling, achieving low complexity and high accuracy across diverse state-space models and noise environment. Experimental validation demonstrates CKFNet's superior robustness and precision over CKF and KalmanNet, particularly in high-noise regimes.

## II. SYSTEM MODEL AND CUBATURE KALMAN FILTER

### A. State Space model

Consider the nonlinear state space model of the discrete system given by

$$\mathbf{x}_i = \mathbf{f}(\mathbf{x}_{i-1}) + \mathbf{w}_{i-1} \qquad (1)$$

This work was partially supported by National Natural Science Foundation of China (grant: 62171388, 61871461, 61571374). (*Corresponding author: Haiquan Zhao*).

Jinhui Hu (e-mail: jhhu_swjtu@126.com), Haiquan Zhao (e-mail: hqzhao_swjtu@126.com) and Yi Peng (e-mail: pengyi1007@163.com) are with the Key Laboratory of Magnetic Suspension Technology and Maglev Vehicle, Ministry of Education, School of Electrical Engineering Southwest Jiaotong University Chengdu, China.

$$\mathbf{z}_i = \mathbf{h}(\mathbf{x}_i) + \mathbf{v}_i \tag{2}$$

where $\mathbf{x}_i \in \mathbb{R}^{n \times 1}$, $\mathbf{z}_i \in \mathbb{R}^{m \times 1}$ represent the *n*-dimensional state vector and *m*-dimensional observation vector, respectively. $\mathbf{w}_i$ and $\mathbf{v}_i$ are the uncorrelated state noise and measurement noise, respectively, and the corresponding noise covariance matrices are $\mathbf{W}_i$ and $\mathbf{V}_i$.

*B. CKF*

The CKF is an effective state estimation algorithm proposed for nonlinear systems, which implements state estimation in two steps, predicting and updating.

1) Predicting

Firstly, the cubature points $\chi_{k,i-1}$ need to be calculated

$$\chi_{k,i-1} = \mathbf{S}_{i-1}\alpha_k + \mathbf{x}_{i-1} \tag{3}$$

where $\alpha_k = \sqrt{n}\delta_k$, $\delta_k$ denotes the *k*-th column of the *n*-dimensional unit matrix, $k = 1, 2, \ldots, n$. $\mathbf{S}_{i-1}$ is the state covariance matrix $\mathbf{P}_{i-1}$ of the algorithm obtained by Cholesky decomposition. Then computing

$$\xi_{k,i|i-1} = \mathbf{f}(\chi_{k,i-1}) \tag{4}$$

$$\mathbf{x}_{i|i-1} = \sum_{k}^{2n} \omega_k \xi_{k,i|i-1} \tag{5}$$

$$\mathbf{P}_{i|i-1} = \omega_k \sum_{k}^{2n} \overline{\xi}_{k,i|i-1} \overline{\xi}_{k,i|i-1}^T + \mathbf{W}_{i-1} \tag{6}$$

where $\overline{\xi}_{k,i|i-1} = \xi_{k,i|i-1} - \mathbf{x}_{i|i-1}$, $\omega_k$ denotes the weights, which usually take $1/2n$.

2) Updating

Calculate the cubature point by

$$\chi_{k,i|i-1} = \mathbf{S}_{i|i-1}\alpha_k + \mathbf{x}_{i|i-1} \tag{7}$$

$$\vartheta_{k,i|i-1} = \mathbf{h}(\chi_{k,i-1}) \tag{8}$$

$$\hat{\mathbf{z}}_{i|i-1} = \omega_k \sum_{k}^{2n} \vartheta_{k,i|i-1} \tag{9}$$

$$\mathbf{P}_{\mathbf{zz},i|i-1} = \omega_k \sum_{k}^{2n} \overline{\vartheta}_{k,i|i-1} \overline{\vartheta}_{k,i|i-1}^T + \mathbf{V}_{i-1} \tag{10}$$

$$\mathbf{P}_{\mathbf{xz},i|i-1} = \omega_k \sum_{k}^{2n} \overline{\vartheta}_{k,i|i-1} \overline{\xi}_{k,i|i-1}^T \tag{11}$$

where $\overline{\vartheta}_{k,i|i-1} = \vartheta_{k,i|i-1} - \hat{\mathbf{z}}_{i|i-1}$. Then computing

$$\mathbf{x}_{i|i} = \mathbf{x}_{i|i-1} + \mathbf{K}_i(\mathbf{z}_i - \hat{\mathbf{z}}_{i|i-1}) \tag{12}$$

$$\mathbf{P}_{i|i} = \mathbf{P}_{i|i-1} - \mathbf{K}_i \mathbf{P}_{\mathbf{zz},i|i-1} \mathbf{K}_i^T \tag{13}$$

$$\mathbf{K}_i = \mathbf{P}_{\mathbf{xz},i|i-1} \mathbf{P}_{\mathbf{zz},i|i-1}^{-1} \tag{14}$$

## III. CKFNET

In this section, we detail the specific architecture of CKFNet, an efficient network structure that builds upon the foundation of the CKF, enhancing interpretability and enabling precise learning of uncertain information within the state estimation process. A key advantage of CKFNet is its use of neural networks to learn the distribution of cubature points, thereby significantly mitigating the impact of cumulative errors from iterative processes on the algorithm's performance.

*A. High Level Architecture*

In this subsection, we introduce CKFNet, a novel network architecture. While its design philosophy is rooted in the approach of KalmanNet, CKFNet goes beyond by not only preserving the cubature point characteristic of CKF in the a priori state estimation but also by leveraging the learning capabilities of neural networks to effectively learn the distribution of these cubature points. This enhancement enables the filter to efficiently mitigate the cumulative errors that arise in the prediction phase.

In CKFNet, we first designing a method for learning the state noise covariance matrix $\mathbf{W}_i$ of the state estimates and the weights of the cubature points using an RNN network at the prediction step of the CKF. In this part the difference between the priori state estimate and the posteriori state estimate at the previous moment is used to learn $\mathbf{W}_i$, $\omega_i$, and $\mathbf{S}_{i|i-1}$ to obtain a relatively ideal distribution of cubature points and weights. Finally, the priori state estimates are computed using (5).

Then at the update step, we similarly design an RNN network to learn the Kalman gain matrix of the CKF. It is similar to the CKF's Kalman gain calculation step, but uses an RNN instead of the step that may contain inaccurate information. The various types of information obtained from the learning of the prediction step are used as part of the input, and the effect of measurement residuals is additionally considered to further learn the Kalman gain matrix. We summarize the structure for CKFNet as shown in Fig. 1.

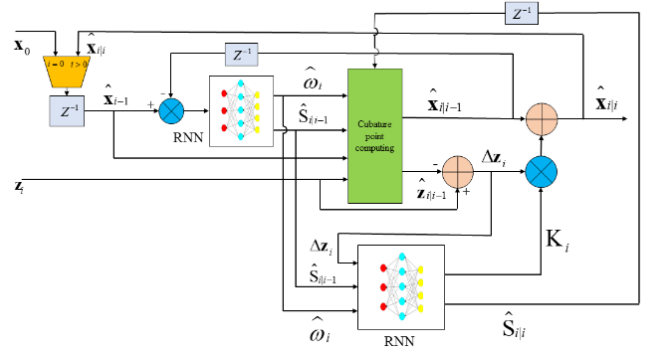

Fig. 1. CKFNet block diagram.

*B. Neural Network Architecture*

In CKFNet's prediction phase, dual GRUs are deployed to refine a priori state estimation–leveraging their temporal modeling efficiency and gated mechanism for adaptive noise characterization. The first GRU dynamically learns the priori covariance matrix, aligning cubature point distributions with ideal posterior approximations through sequential learning capabilities, while the second GRU optimizes cubature point weights in real-time by capturing spatial noise dependencies via its reset/update gates. This dual mechanism overcomes traditional CKF's fixed parametric constraints through GRU's trainable gating, ensuring precise propagation. A subsequent GRU fusion module integrates representations from both GRUs, utilizing their parameter efficiency and vanishing gradient resilience to execute filtering updates. This layered design reflects the strengths of GRU: GRUs enable dynamic uncertainty quantification through learnable covariance propagation, surpassing static model limitations.

In addition, in the update phase, the same two GRU networks are applied to learn the Kalman gain, where the first GRU network is used to learn the mutual covariance matrix in the CKF structure, and the second one is used to learn the measurement error covariance matrix, and it is important to note that both covariance matrices are involved in the learning of the measurement noise distributions, and therefore, the main role of the RNNs here is to learn the unknown measurement noise distribution. Unlike the linearized gain computation of EKF, CKFNet learns the state-measurement cross-covariance matrix and measurement covariance matrix directly through the GRU network and derives Kalman gains based on data-driven adaptive derivation. CKFNet encodes state residuals, innovations, and noise statistics as temporal features, using GRU gates to learn covariance dynamics, enabling explicit modeling and robust estimation of noise distributions in nonlinear systems

The learning of each of the above steps requires the features of the noise distribution during the statistical state estimation process, and therefore features need to be constructed to be used as the features of the GRU, and we use the following signals

F1: Innovation error: $\Delta \hat{\mathbf{z}}_i = \mathbf{z}_i - \hat{\mathbf{z}}_{i|i-1}$ reflects the difference between the observed and predicted values.

F2: Observation error: $\Delta \tilde{\mathbf{z}}_i = \mathbf{z}_i - \mathbf{z}_{i-1}$ quantifies the deviation between state prediction and update.

F3: Forward evolution error: $\Delta \mathbf{x}_i = \mathbf{x}_{i|i} - \mathbf{x}_{i|i-1}$ quantify deviations between state predictions and updates.

F4: Forward updating error: $\Delta \mathbf{x}_i = \mathbf{x}_{i|i} - \mathbf{x}_{i-1|i-1}$ describe the time continuity of state estimation.

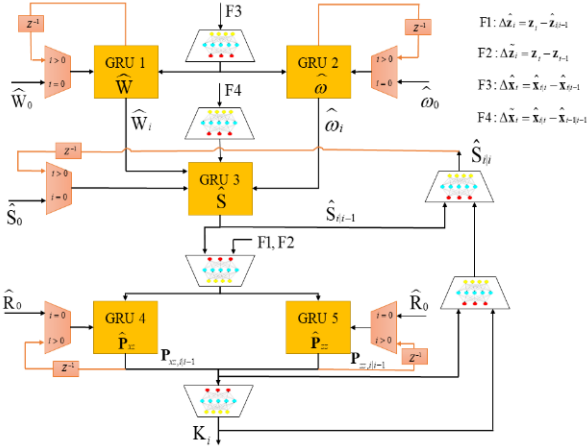

Fig. 2. CKFNet RNN block diagram

The aforementioned features are instrumental in training the overall network architecture. Specifically, features F1 and F2 are predominantly employed during the update phase of CKFNet, where they are crucial for learning the distribution of observation noise. On the other hand, features F3 and F4 are primarily utilized in the prediction step to grasp the distribution of process noise and the behavior of cubature points. The architecture of the proposed network and the detailed configuration of the GRU are depicted in Fig. 2.

### C. Training Algorithm

The CKFNet's training occurs in two sequential stages. First, cubature point distribution and weights are optimized during prediction. These results then update posterior estimates in the next phase, yielding the Kalman gain matrix for state refinement. During training, time-based backpropagation enables gradient learning. Model performance is quantified by a squared-error loss function:

$$\iota = \left\| \mathbf{x}_i - \mathbf{x}_{i|i} \right\|^2 \quad (15)$$

where $\mathbf{x}_i$ refers to the real state and $\mathbf{x}_{i|i}$ refers to the estimated state, the loss function is also used for the performance evaluation of CKF. Next, let $\Xi$ denote the trainable parameters of the RNN and construct a $\ell_2$ regularized mean square error (MSE) loss measure using $\lambda$ as the regularization factor

$$\ell_2(\Xi) = \frac{1}{T_s} \sum_{i=1}^{T_s} \left\| \mathbf{x}_i\left(\mathbf{z}_i^s; \Xi\right) - \mathbf{x}_i^s \right\|^2 + \lambda \left\| \Xi \right\|^2 \quad (16)$$

where the regularization factor $\lambda$ is derived from Bayesian principles. Under the Bayesian framework [30], the posterior distribution of parameters $\theta$ given state estimates $\mathbf{x}$ is

$$p(\theta | \mathbf{x}) \propto p(\mathbf{x} | \theta) p(\theta) \quad (17)$$

where $p(\mathbf{x} | \theta)$ is the Gaussian likelihood: $\exp\left(-\frac{1}{2}\sum_{k=1}^{T} \| \mathbf{x}_k - \hat{\mathbf{x}}_k \|^2\right)$. $p(\theta)$ is the prior distribution of parameters, assumed Gaussian: $\theta \sim N(0, \sigma_\theta^2 I)$. Maximizing the log-posterior is equivalent to minimizing

$$\mathcal{L}(\theta) = \sum_{k=1}^{T} \| \mathbf{x}_k - \hat{\mathbf{x}}_k \|^2 + \frac{1}{\sigma_\theta^2} \| \theta \|^2 \quad (18)$$

Comparing with Eq. (13), we have $\lambda = 1/\sigma_\theta^2$. Setting $\sigma_\theta^2 = 10^4$ yields $\lambda = 0.0001$, allowing sufficient flexibility to learn complex noise dynamics while preventing overfitting.

### IV. SIMULATION RESULT

This section validates CKFNet's performance through numerical simulations. We compare CKFNet with conventional CKF and KalmanNet across diverse scenarios using an automotive land navigation case study. Algorithm performance is evaluated via average MSE (AMSE):

$$\text{AMSE} = \frac{1}{N} \sum_{i=1}^{N} \text{MSE}(i) \quad (19)$$

### A. Linear Land Vehicle Navigation System

We give an example of a navigation system to verify the tracking capability of CKFNet in a navigation system. Firstly, the case of a linear system, the state space model of the system is as follows

$$\mathbf{x}_i = \mathbf{F}\mathbf{x}_{i-1} + \mathbf{w}_{i-1} \quad (20)$$
$$\mathbf{z}_i = \mathbf{H}\mathbf{x}_i + \mathbf{v}_i \quad (21)$$

where

$$\mathbf{F} = \begin{bmatrix} 1 & 0 & \Delta T & 0 \\ 0 & 1 & 0 & \Delta T \\ 0 & 0 & 1 & 0 \\ 0 & 0 & 0 & 1 \end{bmatrix} \quad (22)$$

$$\mathbf{H} = \begin{bmatrix} 1 & 0 & 0 & 0 \\ 0 & 1 & 0 & 0 \\ 0 & 0 & 1 & 0 \\ 0 & 0 & 0 & 1 \end{bmatrix} \quad (23)$$

where $\mathbf{x}_i = \begin{bmatrix} \mathbf{x}_{1,i} & \mathbf{x}_{2,i} & \mathbf{x}_{3,i} & \mathbf{x}_{4,i} \end{bmatrix}$ is the state vector containing the north position, east position, north velocity and east velocity. The noise distribution $\mathbf{w}_{i-1} = N(0,0.1)$ and $\mathbf{v}_i = N(0,0.1)$. The initial parameters are set as $\mathbf{x}_0 = \begin{bmatrix} 0 & 0 & 0 & 0 \end{bmatrix}$, $\mathbf{x}_{0|0} = \begin{bmatrix} 0 & 0 & 0 & 0 \end{bmatrix}$ and $\mathbf{P}_{0|0} = \mathbf{I}_n$. The initial noise covariance for the CKF is set to be $\mathbf{W}_{i-1} = 0.1\mathbf{I}_n$ and $\mathbf{V}_i = 0.1\mathbf{I}_n$. In order to verify CKFNet's ability to learn the observation equations when the observation information is incomplete, we test the performance of the network in this subsection for two cases, with full observation information and with only partial observation information.

**Parameter confirmation:** To evaluate the impact of the number of GRU layers on model performance, we compared the AMSE of CKFNet using 64-layer, 128-layer, and 256-layer GRUs. As shown in Table 1, the 128-layer GRU achieved the lowest AMSE under these conditions. Consequently, we adopted the 128-layer GRU configuration for subsequent experiments.

Table I. AMSE under different layers of GRU

| layers of GRU | 64 | 128 | 256 |
|---|---|---|---|
| AMSE | 0.7945 | 0.7841 | 0.8566 |

**Algorithm Scalability Validation:** CKFNet's generalization is evaluated via multi-stage experiments using 100-timestep training data and extended test horizons (100/120/150/180 timesteps). As shown in fig. 3, CKFNet achieves significant AMSE improvements over KalmanNet while maintaining estimation accuracy comparable to conventional CKF, even with 50% longer test sequences. Results confirm CKFNet's ability to extrapolate learned noise features to unseen scenarios, ensuring stable predictions in extended temporal domains through adaptive learning.

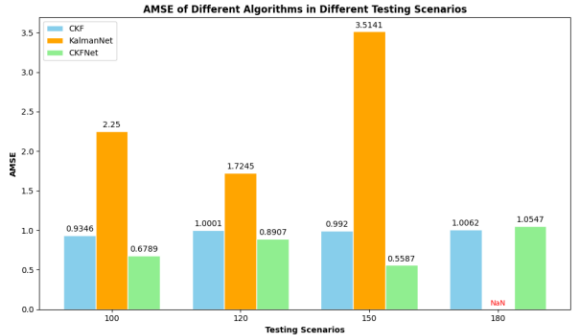

Fig. 3. AMSE of Different Algorithms in Different Testing Scenarios

### B. Nonlinear Land Vehicle Navigation System

Building on the existing state-space framework, we preserve structural continuity in state transitions while introducing nonlinear measurement parametrization. This enables controlled evaluation of observation model nonlinearity impacts via simulations. The state function retains original transition equations, whereas the measurement function adopts a nonlinear observation operator formalized as:

$$\mathbf{z}_i = \begin{bmatrix} -\mathbf{x}_i(1) - \mathbf{x}_i(3) \\ -\mathbf{x}_i(2) - \mathbf{x}_i(4) \\ \sqrt{\mathbf{x}_i^2(1) + \mathbf{x}_i^2(2)} \\ \arctan\left(\sqrt{\dfrac{\mathbf{x}_i(2) - 100}{\mathbf{x}_i(1) - 100}}\right) \end{bmatrix} + \mathbf{v}_i \quad (24)$$

To enhance training data diversity and improve algorithmic robustness, we introduce bounded perturbations to the noise covariance matrix during training. Specifically, we apply a disturbance of ±20% to each eigenvalue of the covariance matrix, thereby augmenting the training data. When evaluating performance under varying noise covariance magnitudes, Fig. 4 shows CKFNet consistently demonstrates advantages across Gaussian noise regimes. This performance gain, achieved through optimized cubature point utilization, translates to significant error reduction. By integrating the geometric principles of the CKF with deep learning, CKFNet establish es dual theoretical and practical advantages in nonlinear stochastic systems, optimally balancing precision and adaptability.

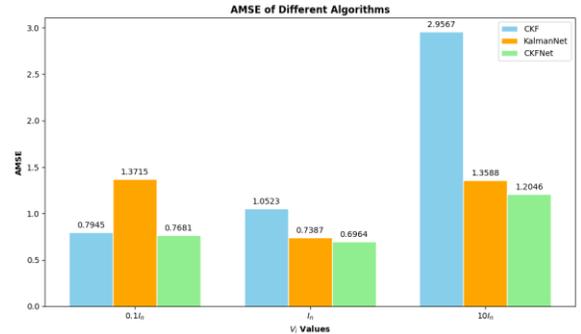

Fig. 4. AMSE of Different Algorithms in different noises

### C. Test time assessment

In this section, we compare the testing times of CKFNet and the other algorithms. Table II compares testing times across algorithms. CKFNet incurs higher computational cost than CKF and KalmanNet due to its enhanced architecture. This complexity enables more accurate noise distribution learning, justifying the trade-off.

Table II. Testing time of Different Algorithms

| Algorithms | CKF | KalmanNet | CKFNet |
|---|---|---|---|
| Testing time/s | 0.8141 | 0.6341 | 0.9929 |

## V. CONCLUSION

This work proposes CKFNet, a hybrid architecture combining RNN-driven learning with CKF's geometric principles for nonlinear state estimation. By embedding RNN modules within the CKF prediction-update cycle, it dynamically models stochastic uncertainties while preserving the mathematical guarantees of cubature point distributions through constrained learning. The framework maintains CKF's interpretability through adaptive weight optimization, enhancing numerical precision without sacrificing analytical tractability. Experiments validate CKFNet's superior performance, confirming its theoretical benefits for nonlinear estimation tasks through neuro-symbolic integration.

## REFERENCES

[1] M. B. Do Coutto Filho and J. C. Stacchini de Souza, "Forecasting-Aided State Estimation—Part I: Panorama," IEEE Transactions on Power Systems, vol. 24, no. 4, pp. 1667-1677, Nov. 2009.

[2] M. Wang, J. Cui, Y. Huang, W. Wu and X. Du, "Schmidt ST-EKF for Autonomous Land Vehicle SINS/ODO/LDV Integrated Navigation," IEEE Transactions on Instrumentation and Measurement, vol. 70, pp. 1-9, 2021.

[3] A. Shahraki, A. Taherkordi, Ø. Haugen and F. Eliassen, "A Survey and Future Directions on Clustering: From WSNs to IoT and Modern Networking Paradigms," IEEE Transactions on Network and Service Management, vol. 18, no. 2, pp. 2242-2274, June 2021.

[4] B. Chen, X. Liu, H. Zhao, Jose C. Principe, "Maximum correntropy Kalman filter", Automatica, Vol. 76, pp. 70-77, 2017.

[5] Y. Tao and S. S. -T. Yau, "Outlier-Robust Iterative Extended Kalman Filtering", IEEE Signal Processing Letters, Vol. 30, pp. 743-747, 2023.

[6] X. Liu, L. Li, L. Zhen, T. Fernando, and H. H. C. Iu, "Stochastic stability condition for the extended Kalman filter with intermittent observations", IEEE Transactions on Circuits and Systems II: Express Briefs, Vol. 64, no. 3, pp. 334–338, Mar. 2017.

[7] I. Arasaratnam and S. Haykin, "Cubature Kalman Filters," IEEE Transactions on Automatic Control, vol. 54, no. 6, pp. 1254-1269, June 2009.

[8] L. Dang, B. Chen, Y. Huang, Y. Zhang and H. Zhao, "Cubature Kalman Filter Under Minimum Error Entropy with Fiducial Points for INS/GPS Integration," IEEE/CAA Journal of Automatica Sinica, vol. 9, no. 3, pp. 450-465, March 2022.

[9] G. Revach, N. Shlezinger, X. Ni, A. L. Escoriza, R. J. G. van Sloun and Y. C. Eldar, "KalmanNet: Neural Network Aided Kalman Filtering for Partially Known Dynamics," IEEE Transactions on Signal Processing, vol. 70, pp. 1532-1547, 2022.

[10] H. Zhao, B. Tian, B. Chen, "Robust stable iterated unscented Kalman filter based on maximum Correntropy criterion, " Automatica, vol. 142, 2022.

[11] B. Chen, L. Dang, Y. Gu, N. Zheng and J. C. Príncipe, "Minimum Error Entropy Kalman Filter," IEEE Transactions on Systems, Man, and Cybernetics: Systems, vol. 51, no. 9, pp. 5819-5829, Sept. 2021.

[12] S. Gao, G. Hu, and Y. Zhong, "Windowing and random weighting-based adaptive Unscented Kalman filter, " International Journal of Adaptive Control and Signal Processing, vol. 29, no. 2, pp. 201–223, 2015.

[13] X. Liu, X. Liu, Y. Yang, Y. Guo and W. Zhang, "Variational Bayesian-Based Robust Cubature Kalman Filter With Application on SINS/GPS Integrated Navigation System," IEEE Sensors Journal, vol. 22, no. 1, pp. 489-500, 1 Jan.1, 2022.

[14] Y. Huang, Y. Zhang, Z. Wu, N. Li and J. Chambers, "A Novel Adaptive Kalman Filter With Inaccurate Process and Measurement Noise Covariance Matrices," IEEE Transactions on Automatic Control, vol. 63, no. 2, pp. 594-601, Feb. 2018.

[15] D. Ćetenović, J. Zhao, V. Levi, Y. Liu and V. Terzija, "Variational Bayesian Unscented Kalman Filter for Active Distribution System State Estimation," IEEE Transactions on Power Systems.

[16] B. Chen, L. Dang, N. Zheng, Jose C. Principe, Kalman Filtering Under Information Theoretic Criteria, Xi'an, China, 2023.

[17] Cho, K., Van Merriënboer, B., Gulcehre, C., Bahdanau, D., Bougares, F., Schwenk, H., & Bengio, Y, "Learning Phrase Representations using RNN Encoder–Decoder for Statistical Machine Translation," Proceedings of the 2014 Conference on Empirical Methods in Natural Language Processing (EMNLP), (2014)

[18] Hochreiter S, Schmidhuber J. "Long short-term memory," Neural Computation, 1997, 9(8): 1735-1780.

[19] M. Zaheer, A. Ahmed, and A. J. Smola, "Latent LSTM allocation: Joint clustering and non-linear dynamic modeling of sequence data," Proc. Int. Conf. Mach. Learn., 2017, pp. 3967–3976.

[20] S. S. Rangapuram, M. W. Seeger, J. Gasthaus, L. Stella, Y. Wang, and T. Januschowski, "Deep state space models for time series forecasting," in Proc. Adv. Neural Inf. Process. Syst., vol. 31, 2018, pp. 1–10.

[21] B. Millidge, A. Tschantz, A. Seth, and C. Buckley, "Neural Kalman filtering," 2021, arXiv:2102.10021.

[22] S. Jouaber, S. Bonnabel, S. Velasco-Forero, and M. Pilte, "NNAKF: A neural network adapted Kalman filter for target tracking," in Proc. IEEE Int. Conf. Acoust., Speech Signal Process. (ICASSP), 2021, pp. 4075–4079.

[23] P. Becker, H. Pandya, G. Gebhardt, C. Zhao, C. J. Taylor, and G. Neumann, "Recurrent Kalman networks: Factorized inference in high-dimensional deep feature spaces," Proc. Int. Conf. Mach. Learn., PMLR, 2019, pp. 544–552.

[24] A. Vaswani et al., "Attention is all you need," in Adv. Neural Inf. Process. Syst., 2017, pp. 5998–6008. [Online]. Available: https://proceedings.neurips.cc/paper/2017/hash/3f5ee243547dee91fbd053c1c4a845aaAbstract.html

[25] H. K. Aggarwal, M. P. Mani, and M. Jacob, "MoDL: Model-based deep learning architecture for inverse problems," IEEE Trans. Med. Imag., vol. 38, no. 2, pp. 394–405, Feb. 2019.

[26] N. Shlezinger, J. Whang, Y. C. Eldar, and A. G. Dimakis, "Model-based deep learning," 2020, arXiv:2012.08405.

[27] R. G. Krishnan, U. Shalit, and D. Sontag, "Deep Kalman filters," 2015, arXiv:1511.05121.

[28] I. Buchnik, G. Sagi, N. Leinwand, Y. Loya, N. Shlezinger and T. Routtenberg, "GSP-KalmanNet: Tracking Graph Signals via Neural-Aided Kalman Filtering," IEEE Transactions on Signal Processing, vol. 72, pp. 3700-3716, 2024.

[29] X. Ni, G. Revach and N. Shlezinger, "Adaptive Kalmannet: Data-Driven Kalman Filter with Fast Adaptation," ICASSP 2024 - 2024 IEEE International Conference on Acoustics, Speech and Signal Processing (ICASSP), Seoul, Korea, Republic of, 2024, pp. 5970-5974.

[30] Bishop, C. M, "Pattern Recognition and Machine Learning," 2006, Springer.